\documentclass{bmcart}

\usepackage{graphicx}
\usepackage{amsthm,amsmath}

\usepackage{cite}

\usepackage[utf8]{inputenc} 

\begin{document}

\begin{frontmatter}

\begin{fmbox}
\dochead{Research}

\title{Ancilla-driven blind quantum computation for clients with different quantum capabilities}

\author[
addressref={aff1},
]{\fnm{Qunfeng} \snm{Dai}}
\author[
addressref={aff2},
]{\fnm{Junyu} \snm{Quan}}
\author[
addressref={aff3},
]{\fnm{Xiaoping} \snm{Lou}}
\author[
addressref={aff1},                   
corref={aff1},                       
email={liqin@xtu.edu.cn}
]{\fnm{Qin} \snm{Li}}

\address[id=aff1]{%
	\orgdiv{School of Computer Science},
	\orgname{Xiangtan University},
	\city{Xiangtan},
	\cny{China}
}
\address[id=aff2]{%
	\orgdiv{School of Mathematics and Computational Science},
	\orgname{Xiangtan University},
	\city{Xiangtan},
	\cny{China}
}
\address[id=aff3]{%
	\orgdiv{College of Information Science and Engineering},
	\orgname{Hunan Normal University},
	\city{Changsha},
	\cny{China}
}

\end{fmbox}


\begin{abstractbox}

\begin{abstract} 
Blind quantum computation (BQC) allows a client with limited quantum power to delegate his quantum computational task to a powerful server and still keep his input, output, and algorithm private. There are mainly two kinds of models about BQC, namely circuit-based and measurement-based models. In addition, a hybrid model called ancilla-driven universal blind quantum computing (ADBQC) was proposed by combining the properties of both circuit-based and measurement-based models, where all unitary operations on the register qubits can be realized with the aid of single ancillae coupled to the register qubits. However, in the ADBQC model, the quantum capability of the client is strictly limited to preparing single qubits. If a client can only perform single-qubit measurements or a few simple quantum gates, he may also want to delegate his computation to a remote server via ADBQC. This paper solves the problem and extends the existing model by proposing two types of ADBQC protocols for clients with different quantum capabilities, such as performing single-qubit measurements or single-qubit gates. Furthermore, in the proposed two ADBQC protocols, clients can detect whether servers are honest or not with a high probability by using corresponding verifiable techniques.
\end{abstract}




\end{abstractbox}
%

\end{frontmatter}




\section{Introduction}
The implementation of quantum computing is generally based on circuit-based model \cite{deutsch1989quantum,broadbent2015delegating,zhang2018single} and measurement-based model \cite{raussendorf2001one,takeuchi2019quantum,dunjko2012blind,morimae2012blind,morimae2013blind,morimae2013secure,li2014triple,sheng2015deterministic,miller2016hierarchy}. In the circuit-based model, quantum computing is realized by directly acting single-qubit or multi-qubit gates on the qubits in quantum registers. In contrast, the measurement-based model is implemented by performing adaptive single-qubit measurements on a highly entangled resource state. Since the two models can simulate each other, they are computationally equivalent. Each model has its own advantages and disadvantages and which one is chosen mainly depends on the physical system and the quantum devices of the user. 

In 2010, a mixture of the two models, called ancilla-driven quantum computation (ADQC), was proposed by Anders et al.\cite{anders2010ancilla}, where qubits are stored in quantum registers like the circuit-based model, whereas the operations on the register are performed by measuring an ancilla attached to the register in different bases similar to the measurement-based model. The main feature of ADQC is that the ancilla qubit is coupled to various qubits of register through a fixed two-qubit entanglement operator $(H\otimes H) CZ$, and only the ancilla qubit is initialized and measured. Due to the entanglement effect of the register and ancilla, arbitrary quantum operations on qubits of the register can be realized by performing suitable measurements on the ancilla. ADQC has excellent advantages in some physical systems where register qubits with long decoherence time are difficult to operate, while relatively short-lived ancilla qubits are easier to control and can be prepared and measured quickly, such as neutral atoms in optical lattices \cite{jaksch1999entanglement}, cavity QED superconducting qubits \cite{majer2007coupling}, and aluminum ions in optics \cite{chou2010frequency,effecton}. Besides, ADQC can simulate any positive operator valued measurement (POVM) on register qubits by accessing a fully controlled ancilla which is attached to the register sequentially. Therefore, it is also useful for experimental systems where their measurements would destroy physical qubits, such as photonic systems.

Although quantum computation has been extensively studied, the physical realization of it is still very challenging. Even if quantum computers become available, they are likely to be owned by only a handful of centers around the world much like today's supercomputer rental system. Clients who want to utilize these quantum resources can only delegate their computational tasks to the organizations that own quantum computers. The burdens of clients are greatly reduced in such a delegated quantum computing model, but their privacy is seriously threatened. Fortunately, some quantum cryptographic techniques, such as quantum  key distribution \cite{price2020quantum,quantumkey}, quantum identity authentication \cite{Li2018blind,LI2021253}, and quantum secret sharing \cite{qin2020hierarchical}, can be utilized to protect the privacy of clients.  

Blind quantum computation (BQC) as a combination of quantum computation and quantum cryptography is a kind of delegated quantum computing that can protect private data of clients. It allows a client who only has some simple quantum devices to delegate quantum computing tasks to a powerful quantum server, while keeping the data of the client including input, output, and algorithm hidden from the server.
The first BQC protocol was proposed by Childs based on the circuit model \cite{childs2005secure}, where the client Alice must possess quantum memory, prepare $|0\rangle$, and have the ability to perform $SWAP$ gates. Broadbent, Fitzsimons, and Kashefi proposed the first universal BQC protocol (known as the BFK protocol) \cite{broadbent2009universal}, in which the client only needs to prepare single-qubit states and does not require quantum memory and the ability to perform complex quantum gates. Then Morimae et al. proposed another BQC model \cite{morimae2013blind} in which the client only makes measurements, as in some experimental settings such as quantum optical systems, the measurement of a qubit is much easier than generating a single-qubit state. Since then, a series of BQC protocols were proposed based on these two protocols \cite{gheorghiu2015robustness,hayashi2015verifiable,fitzsimons2017unconditionally,morimae2016measurement,huang2017universal,hayashi2018self,morimae2014verification,takeuchi2018verification,sato2019arbitrable,shan2021multi,ZHANG2019135} and a few proof-of-principle experiments were demonstrated in photonic systems \cite{barz2012demonstration,barz2013experimental}. Recently, Li et al. proposed a new model of BQC where a client only needs to perform several single-qubit gates\cite{li2021blind} and it provides a new research path for BQC.

An ancilla-driven blind quantum computing (ADBQC) protocol was proposed by applying the BQC technology to the ADQC model \cite{sueki2013ancilla}, which realized the ADQC in the way of delegated quantum computing for the first time. After that, another ADQC protocol without performing measurements was proposed to further enrich the field of ADQC \cite{proctor2014minimal}. In ADBQC, it is implemented in a very monolithic way, and clients should generate various single qubits. In fact, it is unrealistic that all users have the same quantum ablitity. As mentioned above, BQC mainly deals with three types of clients. Therefore, it is also necessary to design ADBQC protocols suitable for various clients with different quantum capability, such as performing single-qubit measurements or gates. This paper extends the existing ADBQC model by proposing two ADBQC protocols for another two kinds of clients who only have the ability to perform single-qubit measurements or gates. Moreover, the proposed ADBQC protocols can be verifiable, as clients can easily verify whether the server deviates from the calculation by introducing the trap qubit technology.

The rest of this paper is organized as follows. Section 2 describes the preliminaries, including basic notations and structure of the circuit gadgets needed to realize ADBQC. In section 3, we briefly review a typical ADBQC protocol for the users who prepare single-qubit states  \cite{sueki2013ancilla}. Section 4 presents two ADBQC protocols for another two types of users and analyzes the security and the verifiability of them. The last section gives a conclusion of this paper.

\section{Preliminaries}
In this section, we give a brief introduction to ADBQC. A more detailed description is available in \cite{anders2010ancilla,anders2012ancilla,sueki2013ancilla}. There are two types of qubits in ADBQC: register qubit and ancilla qubit. The role of ancilla qubit is to indirectly control the evolution of the register qubit by performing operations such as single-qubit gates and single-qubit measurements on the ancilla qubit after establishing the entanglement between the ancilla and register qubits. We first review the notations and unitary matrices used in a typical ADBQC protocol \cite{sueki2013ancilla}, then present the structure of circuit gadgets which can be used to simulate $HR_{Z}(\theta)$ and $CZ$ gates. By combining these gadgets, ADQC can be realized blindly in the form of delegated computation. In addition, we refer to the client as Alice and the server as Bob for simplicity.

\subsection{Review of ADBQC in Ref. \cite{anders2012ancilla}}
Let the notation $\{|\pm\rangle\}$ denote $X$ basis measurement and $\{|0\rangle,|1\rangle\}$ denote $Z$ basis measurement. Measurement outcome is represented by $S_{i} \in \{0,1\}$ associated with $\pm$ and the subscript $i$ of $s$ means the $i$-th measurement. Set the state $ |+_{\alpha,\varphi }\rangle=\cos(\frac{\alpha}{2})|0\rangle+e^{i \varphi}\sin(\frac{\alpha}{2})|1\rangle$, the state $ |-_{\alpha,\varphi }\rangle=\cos(\frac{\alpha}{2})|0\rangle-e^{i \varphi}\sin(\frac{\alpha}{2})|1\rangle$, the rotation operator about the $x$ axe
$R_{X}(\theta)=e^{-\frac{i \theta X}{2}}$, and the rotation operator about the $z$ axe $R_{Z}(\theta)=e^{-\frac{i \theta Z}{2}}$. And the Pauli matrices are defined as 
\begin{equation} 
	\label{eq:1}
	 X=\left[
	\begin{array}{cc}
		0&1\\
		1&0\\
	\end{array}
	\right]
	,    	Y=\left[
	\begin{array}{cc}
		0&-i\\
		i&0\\
	\end{array}
	\right]
	,    	Z=\left[
	\begin{array}{cc}
		1&0\\
		0&-1\\
	\end{array}
	\right]  
\end{equation}

ADBQC is performed with the help of different single ancillae, on which single-qubits measurements and 2-qubit entangle operators $\widetilde{E}_{AR}$ determined by the ADBQC scheme chosen by Bob and the computation progress are carried out, where $\widetilde{E}_{AR}=(H\otimes H)CZ$ or $CZ(H\otimes H)$. An ancilla $|+_{\gamma, \delta}\rangle$ is coupled to register qubits with $\widetilde{E}_{AR}$ and then is measured in certain basis
$\{|\pm_{\theta, \phi}\rangle\} $. The back-ation of this measurement on the register qubit can be described by a Kraus operator $K_{\pm}=_{A}\langle\pm_{\theta,\phi}|\widetilde{E}_{AR}|+_{\gamma,\delta}\rangle_{A}$\cite{khaneja2001time}, and $P^{\pm}$=$tr(K^{\pm\dagger}_{R}K^{\pm}_{R})$ are the probabilities of obtaining measurement outcomes $+$ or $-$.

Arbitrary single-qubit gates together with the $CNOT$ gate form the universal set of gates for quantum computation. Two ways are used in ADBQC to carry out arbitrary single-qubit gates, one using $HR_{Z}{(\theta)}$ and the other using $R_{X}(\theta)$ and $R_{Z}(\theta)$, as any unitary operation $U$ can be decomposed as follows: 
\begin{equation} 
		\label{eq:2}
		U=e^{i\alpha}HR_{Z}(0)HR_{Z}(\beta )HR_{Z}(\gamma)HR_{Z}(\delta)\\
		= e^{i\alpha}R_{Z}(\beta )R_{X}(\gamma)R_{Z}(\delta),\\
\end{equation} 
where $\alpha$, $\beta$, $\gamma$, and $\delta$ are real numbers. The matrix of $HR_{Z}(\theta)$, $R_{Z}(\theta)$ and $R_{X}(\theta)$ are presented below.
\begin{equation} 
	\begin{split}
		\label{eq:3}
		HR_{Z}(\theta)=&\frac{1}{\sqrt{2}}\left[
		\begin{array}{cc}
			1&e^{i\theta}\\
			1&-1e^{i\theta}\\
		\end{array}
		\right]
		,   R_{Z}(\theta)=\left[
		\begin{array}{cc}
			1&0\\
			0&e^{i\theta}\\
		\end{array}
		\right], \\
		R_{X}(\theta)=&\left[
		\begin{array}{cc}
			cos\frac{\theta}{2}&-isin\frac{\theta}{2}\\
			-isin\frac{\theta}{2}&	cos\frac{\theta}{2}\\
		\end{array}
		\right]    
	\end{split}
\end{equation}
\\ 

\subsection{Circuit gadgets of ADBQC}
Here we only describes the circuit gadgets for implementing the $CZ$ gate and $HR_{Z}(\theta)$, while gadgets for achieving $R_{X}(\theta)$ and $R_{Z}(\theta)$ are similar. More details can be found in Ref. \cite{sueki2013ancilla}.

\begin{figure}[h!]
	\centering
	\includegraphics[width=1.0\linewidth]{"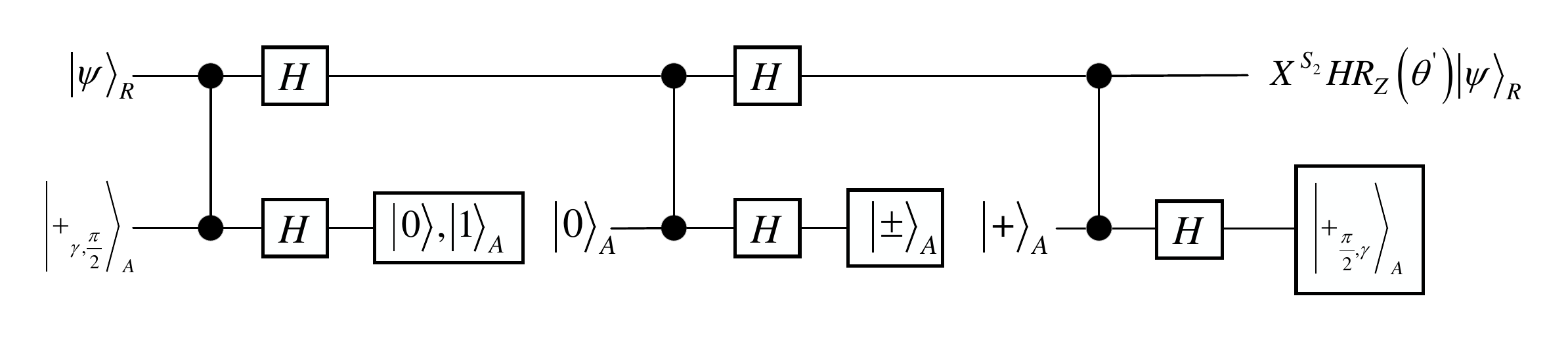"}
	\caption{Schematic structure of a gadget used to achieve the operator $HR_{Z}(\theta')$, where subscripts $R$ and $A$ indicate the register and ancilla qubit, respectively. The state of each ancilla qubit is uniformly and randomly initialized as $|\pm_{\gamma, \frac{\pi}{2}}\rangle$ by Alice and then it is sent to Bob. $\theta'$ is the rotation angle that Alice actually wants to perform in her computation, where $\theta'=-\theta-(-1)^{S_{1}}\gamma$. Alice sends $\theta$ to Bob through the classical channel.}
	\label{fig:simulation-of-hrztheta}
\end{figure}

\begin{figure}[h]
	\centering
	\includegraphics[width=1.0\linewidth]{"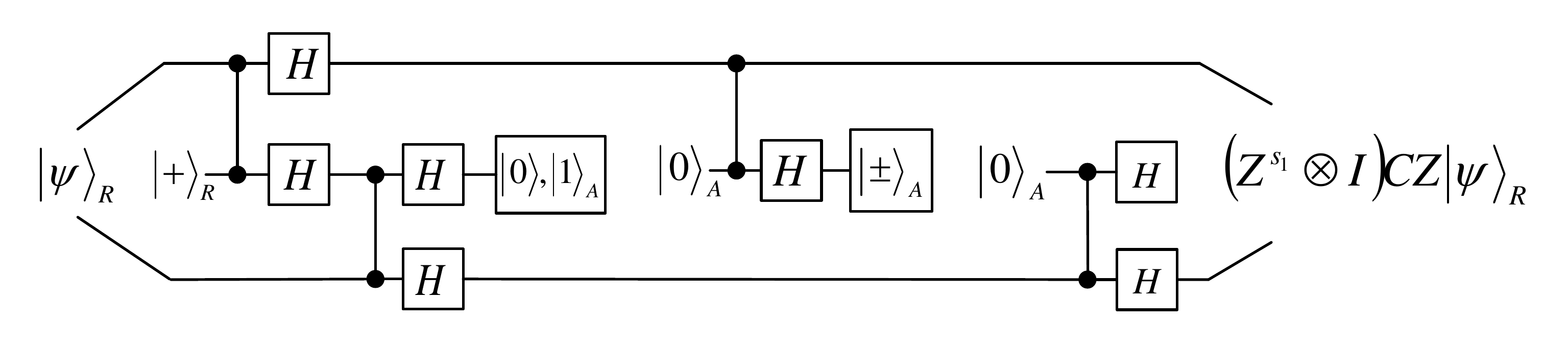"}
	\caption{Schematic structure of a gadget used to achieve the $CZ$ gate. The subscript $A$ stands for ancilla qubits and the subscript $R$ stands for register qubits.}
	\label{fig:simulation-of-cz}
\end{figure}

\begin{figure}[h]
	\centering
	\includegraphics[width=1.0\linewidth]{"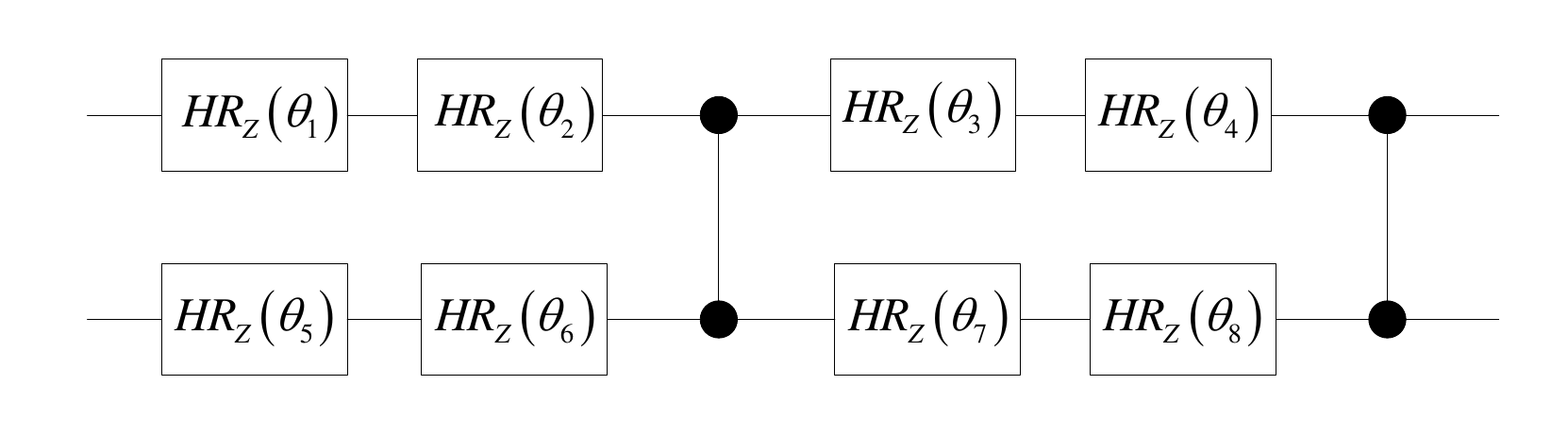"}
	\caption{Schematic structure of a circuit diagram which can be used to simulate the $CNOT$ gate and arbitrary single-qubit gates. The operator $HR_{Z}(\theta)$ in each rectangular box in this circuit diagram is simulated by the gadget in Fig. \ref{fig:simulation-of-hrztheta}, and the $CZ$ operator is implemented by the simulation in Fig. \ref{fig:simulation-of-cz}.}
	\label{fig:a-gate-pattern}
\end{figure}

\begin{figure}[h]
	\centering
	\includegraphics[width=10cm]{"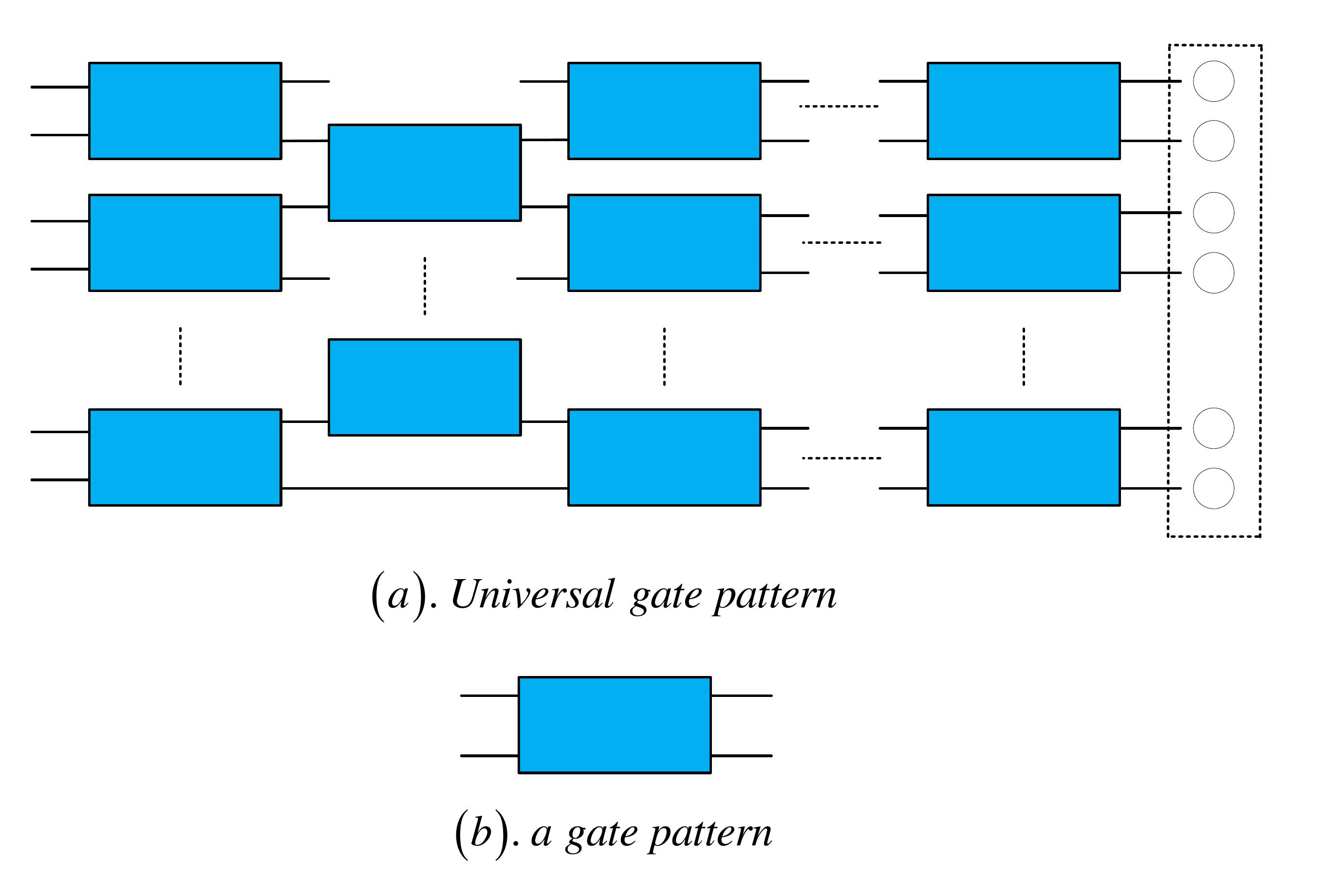"}
	\caption{(Color online) Universal gate pattern. The circuit used to achieve ADBQC is presented here, and the internal circuit in the blue rectangle is shown in Fig.\ref{fig:a-gate-pattern}. The part framed by the dashed line is the output quantum state.}
	\label{fig:universal-gate-pattern}
\end{figure}
In the ADBQC model, a circuit is constructed by Bob and Alice layer-by-layer to execute ADBQC as
shown in Fig. \ref{fig:universal-gate-pattern}(a). Actually, universal BQC can be realized by simulating arbitrary one or two-qubit gates using the universal gate pattern as shown in Fig. \ref{fig:universal-gate-pattern}$(b)$. The universal gate pattern consists of $HR_{Z}(\theta)$ and $CZ$ gates and whether it simulates a one or two-qubit gate is determined by the angle $\theta$ in each $HR_{Z}(\theta)$. Thus, universal ADBQC can be achieved by using the gadgets in Fig. \ref{fig:simulation-of-hrztheta} and Fig. \ref{fig:simulation-of-cz} to implement $HR_{Z}(\theta)$ and the $CZ$ gates.
\section{Review of Sueki et al.'s ADBQC protocol \cite{sueki2013ancilla}}
The ADBQC protocol proposed by Sueki et al. \cite{sueki2013ancilla} is briefly reviewed in this part.
In this protocol, Alice who can only generate single-qubit states can perform ADQC with the help of server Bob while keeping Alice's privacy including input, output, and algorithm perfectly secret. 

Assume the client Alice needs to choose an appropriate circuit gadget to implement computation on register qubits. If Alice wants to execute the operation $HR_{Z}(\theta)$, she needs to prepare and send ancilla $|+_{\gamma,\frac{\pi}{2}}\rangle$ or $|-_{\gamma,\frac{\pi}{2}}\rangle$ to Bob with equal probabilities, where $\gamma$ is chosen randomly by Alice. Then, according to the measurement result sent by Bob, Alice will update the selection of the next measurement angle. However, if Alice needs to achieve the $CZ$ gate, she only needs to receive the classical measurement $S^{'}_{1}$ from Bob.
The specific steps can be briefly described as follows.

(T1) Resource preparation phase:
The circuit scale $N\times M$ required to execute ADBQC is predetermined by Alice and Bob before the protocol as shown in Fig. \ref{fig:universal-gate-pattern} $(a)$, where $N$ is the number of register qubits and $M $ is the depth of the algorithm Alice wants to execute. All the register qubits are initialized in $|0\rangle$. 

(T2) Calculation phase: For every gate pattern in Fig. \ref{fig:universal-gate-pattern} $(b)$, it is composed of the $CZ$ operator and $HR_{Z}(\theta)$ operator. The $CZ$ operation is completed by Bob according to the calculation progress, and Bob only needs to inform Alice of the measurement results $S'_{1}$ through the classic channel. The realization of the $HR_{Z}(\theta)$ operator requires mutual interaction between Alice and Bob and the following four steps: 

(1) Alice chooses an ancilla parameter $\gamma$ randomly and secretly and then sends one of the ancilla qubits $\{|+_{\gamma,\frac{\pi}{2}}\rangle, |-_{\gamma,\frac{\pi}{2}}\rangle\}$ to Bob. Bob couples the ancilla to a register qubit and then measures it in $Z$ basis. After performing the measurement on the ancilla, Bob sends the outcome $s_{1}$ to Alice.

(2) Bob prepares $|0\rangle$ as an ancilla and couples it to the register qubit via the two-qubit unitary $(H\otimes H)CZ$. The purpose of this step is to generate the $H$ gate on the register qubit to offset the redundant $H$ gate generated in the previous step.  

(3) Alice calculates $\theta=-\theta^{'}-(-1)^{S_{1}} (\gamma+r\pi)$ with a random bit $r \in \{0,1\}$ and sends $\theta$ to Bob through a classic channel, where $\theta'$ is the actual rotation angle in the computation. Because Bob does not know the value of $\gamma$, he cannot deduce the real calculated angle through $\theta$. Bob couples the ancilla $|+\rangle$ to the register qubit and measures the coupled ancilla in $\{|+_{\frac{\pi}{2},\theta}\rangle, |-_{\frac{\pi}{2},\theta}\rangle\}$. Then Bob sends the measurement result $S_{2}$ to Alice.

(4) Alice updates the value of $\theta'$ to correct Pauli by-products from the measurements $S_{2}$ and $S^{'}_{1}$ by the method similar to that used in MBQC \cite{broadbent2009universal}.   

(T3) Output phase: At the end of the protocol, Alice will instruct Bob to apply appropriate measurement on each output qubit. Since the output quantum states are encrypted by Pauli-$X$ and $Z$ gates, Bob cannot obtain Alice's actual output from the measurements. Alice simply chooses whether to flip the classical measurement results or not according to the Pauli by-products related to each output qubit. If Bob is honest, Alice will get the correct calculation.

In the reviewed ADBQC protocol \cite{sueki2013ancilla}, only clients with the ability to prepare single quantum states can perform ADBQC. This limits the opportunity for clients with only other fundamental quantum capabilities to participate in ADBQC. Furthermore, the protocol lacks verifiability and Bob could easily mess up the computation to cheat Alice.
\section{The Presented ADBQC protocols }
This section extends the existing ADBQC model for users who prepare single-qubit states in Ref. \cite{sueki2013ancilla} by proposing two ADBQC protocols, called Protocol 1 and Protocol 2, for another two types of users. The proposed protocols both consist of three parts: the resource preparation phase, the computation phase, and the verification phase, where the steps in the resource preparation phase and the verification phase are similar, but the $HR_{Z}(\theta)$ is implemented in a different way in the computation phase. Each of these protocols has a different minimum quantum
capability requirement for the client, such as performing single-qubit measurements in Protocol 1 and performing some single-qubit gates in Protocol 2. In addition, to satisfy verifiability, we must perform measurements directly on the output register qubits rather than using the ADBQC to simulate POVM. Through the trap checking, Alice can verify the computation result with a high probability.
\subsection{The Presented ADBQC protocols in which Alice only makes measurements }
In Protocol 1, Alice who only has the ability to perform Pauli- $X$, $Y$, and $Z$ basis measurements wants to perform ADQC on register qubits with the help of Bob while keeping her own data hidden. The detailed steps of Protocol 1 are given as follows.

(Q1) Resource preparation phase: The circuit size $N\times M$ is determined by Alice and Bob before starting the protocol, where $N$ is the number of register qubits and $M $ is the depth of the algorithm Alice wants to execute. All the register qubits are initialized in $|0\rangle$. Alice chooses $2N/3$ as the number of trap qubits, which is optimal\cite{morimae2014verification}. 

(Q2) Calculation phase: The gate pattern shown in Fig. \ref{fig:a-gate-pattern} is used to perform ADBQC in this protocol.
The gate pattern is composed of the $CZ$ gate and $HR_{Z}(\theta)$.

For the simulation of the $CZ$ gate in each gate pattern in Fig. \ref{eq:2}, the operations are the same as the reviewed Sueki et al.'s protocol. Bob should send the first measurement result $S^{'}$ to Alice for removing the Pauli by-products. 

\begin{figure}[h]
	\centering
	\includegraphics[width=1.0\linewidth]{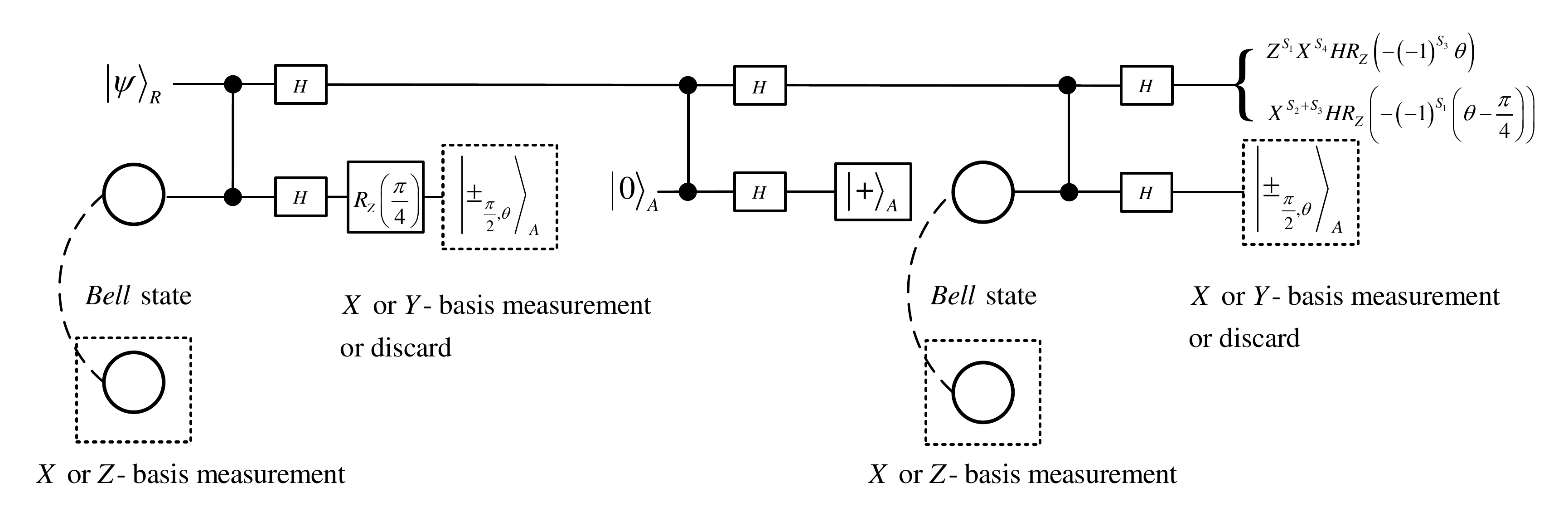}
	\caption{Schematic diagram of achieving the operation $HR_{Z}(\theta)$ in Protocol 1, where angle $\theta$ is restricted to $A \equiv \{0, \frac{2\pi}{4}, \frac{4\pi}{4}, \frac{6\pi}{4}\}$. The implementation of each $HR_{Z}(\theta)$ operation requires the participation of two ancillary qubits. Each ancilla is produced by Alice's $X$ or $Z$-basis measurement on half of the Bell state $\frac{1}{\sqrt{2}}(|00\rangle+|11\rangle)$. In addition, Alice needs to perform X or Y-basis measurements on the ancillary qubits that have been manipulated by Bob to drive the computation on register qubits. Real number $S_{i}$ denotes the i-th measurement result.}
	\label{fig:alicemeasurements}
\end{figure}

For the simulation of the operation $HR_{Z}(\theta)$ in each gate pattern in Fig. \ref{fig:alicemeasurements}, two sets $A \equiv \{0, \frac{2\pi}{4}, \frac{4\pi}{4}, \frac{6\pi}{4}\}$ and  $B \equiv \{\frac{\pi}{4}, \frac{3\pi}{4}, \frac{5\pi}{4}, \frac{7\pi}{4}\}$ are defined. If the measurement angle that Alice needs to perform according to her algorithm is in set $A$, Alice and Bob follow the steps of case $a$. Otherwise, they turn to the case $b$. 

case $a$:
(1) Bob prepares the Bell state $\frac{1}{\sqrt{2}}(|00\rangle+|11\rangle)$ and sends half of it to Alice, keeping the remaining particle as the ancilla. (2) Alice performs an $X$-basis measurement on the particle sent by Bob. After her measurement, Bob has the state $Z^{a}|+\rangle$, where $a \in \{0, 1\}$ is Alice's measurement result. (3) Bob couples the ancilla to the register qubit using the 2-qubit operator $(H\otimes H)CZ$ and then performs the $R_Z(\frac{\pi}{4})$ gate on the coupled ancilla. Later, it is sent to Alice through quantum channel. (4)Alice performs the $X$ or $Y$-basis measurement on the ancilla according to her algorithm. (5) Bob prepares $|0\rangle$ and performs the coupling operator $(H\otimes H)CZ$ on $|0\rangle$ and the register qubit. (6) Bob prepares Bell state $\frac{1}{\sqrt{2}}(|00\rangle+|11\rangle)$ again, sends Alice half of it and keeps the other particle as the ancilla. (7) Alice measures the particle sent by Bob in the $Z$ basis. After her measurement, Bob has the state $X^{b}|0\rangle$, where $b \in \{0, 1\}$ is Alice's measurement result. (8) Bob performs the 2-qubit gate $(H\otimes H)CZ$ on the ancilla and register qubit, then sends the ancilla to Alice. (9) Alice discards the ancilla sent by Bob directly. At last, Alice implements the  operation  $Z^{S_{1}}X^{S_{4}}HR_{Z}(-(-1)^{S_{3}}\theta)$.

case $b$: 
(1) Bob prepares the Bell state $\frac{1}{\sqrt{2}}(|00\rangle+|11\rangle)$ and sends one of two qubits to Alice while keeping the remaining one as the ancilla. (2) Different from that in case $a$, Alice performs an $Z$-basis measurement on the particle sent by Bob here. According to Alice measurement result $a \in \{0, 1\}$, Bob has the state $X^{a}|0\rangle$. (3) Bob performs the 2-qubit gate $(H\otimes H)CZ$ on the ancilla and register qubits, then performs the gate $R_{Z}(\frac{\pi}{4})$ on the ancilla. Later, Bob sends it to Alice via the quantum channel. (4) Alice discards the ancilla sent by Bob. (5) Bob prepares $|0\rangle$ and performs the coupling operator $(H\otimes H)CZ$ on the register qubit and $|0\rangle$. (6) Bob prepares the Bell state $\frac{1}{\sqrt{2}}(|00\rangle+|11\rangle)$ again, sends Alice half of it and keeps the other particle as ancilla. (7) Alice measures the particle sent by Bob in the $X$ basis. After her measurement, Bob has the state $Z^{b}|+\rangle$, where $b \in \{0, 1\}$ is Alice's measurement result. (8) Bob couples the ancilla to the register qubit using the $(H\otimes H)CZ$ gate and sends the coupled ancilla to Alice. (9) Alice measures the ancilla sent by Bob in the $X$ or $Y$ basis according to the algorithm. By these steps, Alice achieves the operation $X^{S^{2}+S^{3}}HR_{Z}(-(-1)^{S_{1}}(\theta - \frac{\pi}{4}))$.

(Q3) verification phase: Assume the output state after Q2 is $\sigma_{q}P|\Psi\rangle$ $=$ $\sigma_{q} P(|\zeta\rangle_{r} \otimes |0\rangle_{t}^{\otimes N/3} \otimes |+\rangle_{t}^{\otimes N/3})$, where $\sigma_{q} \equiv \otimes_{j=1}^{N} X_{j}^{x_{j}} Z_{j}^{z_{j}} $ with $x_{j}$ and $z_{j}$ $\in \{0,1\}$ are Pauli by-products similar to the measurement-based quantum computation \cite{broadbent2009universal}, $P$ is an $N$-qubit permutation, and $|\zeta\rangle_{r}$ is a quantum state of $N/3$ qubits, consisting of $|\pm\rangle$, $|0\rangle$, and $|1\rangle$. We denote the qubits with subscript $r$ are the actual output qubits and the qubits with subscript $t$ are trap qubits. Bob sends qubits of state $\sigma_{q}P|\Psi\rangle$ to Alice one by one. Alice measures the qubits in X or Z bases. If the error rate of the trap qubits is acceptable, Alice accepts the results of these computational register qubits. Otherwise, she rejects them.

\emph{Security analysis. } The blindness and verifiability of Protocol 1 are analysed. Furthermore, blindness is divided into algorithmic blindness, input blindness, and output blindness.

\emph{Algorithm blindness of Protocol 1.} In protocol 1, there is only one-way transmission of information from Bob to Alice, Thus Alice's privacy is guaranteed by the no-signaling principle \cite{popescu1994quantum}. Let $A$ be the random variable representing the angle of Alice's measurement. Let $B$ be the random variable representing the type of POVM performed by Bob, and $M_{B}$ is a random variable representing the outcome of Bob's POVM. Let $T$ be the random variable that Bob sends to Alice and it represents the measurement result when simulating the $CZ$ gate. Because of the no-signaling principle, 
\begin{equation}
	\begin{aligned}
		P(M_{B}=m_{B}|A=a,B=b)
		=P(M_{B}=m_{B}|A=a^{'},B=b),
	\end{aligned}
\end{equation}
for all $M_{B}$, $a$, $a^{'}$, and $b$. From Bayes' theorem, there is the relationship between probabilities,
\begin{equation}
	\begin{aligned}
		&P(A = a | B = b, M_{B} = m_{B}, T = t)\\
		=&\frac{P(M_{B} = m_{B}, A = a, B = b, T = t)}{P(B = b, M_{B} = m_{B}, T = t)}\\
		=&\frac{P(M_{B} = m_{B}, A = a, B = b)P(T = t)}{P(B = b, M_{B} = m_{B}, T = t)}\\
		=&\frac{P(M_{B} = m_{B}|A = a, B = b)P(A = a,B = b)P(T = t)}{P(B = b, M_{B} = m_{B},T=t)}\\
		=&\frac{P(M_{B} = m_{B} | A = a^{'}, B = b)P(A = a^{'}, B = b)P(T = t)}{P(B = b, M_{B} = m_{B}, T = t)}\\
		=&P(A = a^{'} | B = b, M_{B} = m_{B}, T = t).
	\end{aligned}
\end{equation}

This means the conditional probability distribution of Alice's computational angles is equal to its priori probability
distribution. So Bob cannot learn anything about Alice's measurement angles, the blind of algorithm of Protocol 1 is guaranteed.

\emph{Input blindness of Protocol 1.} Let the initial state of the computation be the standard state $|0\rangle^{\otimes N}$ and the algorithm for computation part includes the preparation of the input state. It has been shown above that the algorithm of Protocol 1 are blind, so Bob has no way of knowing what input state Alice prepared.

\emph{Output blindness of Protocol 1.}
Let $O$ be the random variable representing Alice's output, $B$ be the random variable that represents the type of POVM carried out by Bob, $M_{B}$ be the random variable representing the outcome of Bob's POVM, and $T$ be the random variable that Bob sends to Alice representing the measurement result when simulating the $CZ$ gate. Because of the no-signaling principle,
\begin{equation}
	\begin{aligned}
		P(M_{B} = m_{B}|0 = o, B = b)
		=P(m_{B} = m_{B}|O = o^{'}, B = b),
	\end{aligned}
\end{equation}
for all $M_{B}$, $o$, $o^{'}$, and $b$. Then, from Bayes' theorem,
\begin{equation}
	\begin{aligned}
		&P(O =o|B=b, M_{B}=m_{B}, T=t)\\
		=&\frac{P(M_{B} = m_{B}, O = o, B = b, T= t)}{P(B=b, m_{B} = m_{B}, T=t)}\\
		=&\frac{P(M_{B} = m_{B}, A = a, B = b)P(T = t)}{P(B = b, M_{B} = m_{B}, T = t)}\\
		=&\frac{P(M_{B} = m_{B}|O = o, B = b)P(O = o,B = b)P(T = t)}{P(B = b, M_{B} = m_{B},T=t)}\\
		=&\frac{P(M_{B} = m_{B} | O = o^{'}, B = b)P(O = o^{'}, B = b)P(T = t)}{P(B = b, M_{B} = m_{B}, T = t)}\\
		=&P(O = o^{'} | B = b, M_{B} = m_{B}, T = t).
	\end{aligned}
\end{equation}
As the conditional probability of Alice's output is equal to it's prior probability, Bob cannot learn anything about the Alice's output.

\emph{Verifiability of Protocol 1.} Assuming that Bob is dishonest and he will not implement some steps of the protocol as required. His general attack is to create a different state $\rho$ insted of $\sigma_{q}|\Psi\rangle$. This attack can be deduced to a random Pauli attack by a completely positive-trace preserving (CPTP) map \cite{morimae2014verification}.\\
$\mathbf{Proof.}$ Our notations follows that of Ref.\cite{morimae2014verification}. We define $\alpha$ as the number of non-trivial Pauli operators acting on the $N$ register qubits in a random Pauli attack, where non-trivial Pauli operators means $X$, $Z$, and $XZ$ operators.  Let $a$, $b$, $c$ be the number of $X$, $Z$, and $XZ$ operators in $\alpha$. Since $\alpha$ $= a + b + c$ $\leq$ 3max($a$,$b$,$c$), we have max($a$,$b$,$c$) $\geq$ $\frac{\alpha}{3}$.

Let $max(a,b,c)$ $=$ $a$. Then, the probability 
that all $X$ operators of $\sigma_{\alpha}$ do not change any trap is$\frac{(N-a)!\prod_{K=0}^{a-1}(\frac{2N}{3}-k)}{(N)!}$ $=$ $(\frac{2}{3})^{a}\frac{\prod_{K=0}^{a-1}(N-\frac{3K}{2})}{\prod_{K=0}^{a-1}(N-K)}$ $\leq$ $(\frac{2}{3})^{a}$ $\leq$ $(\frac{2}{3})^{\alpha/3}$.
We can obtain the same result for max$(a,b,c)$ $=$ $b$. For
max$(a,b,c) = c$, we have the probability $\frac{(N-a)!\prod_{K=0}^{a-1}(\frac{N}{3}-k)}{(N)!}$ $=$ $(\frac{1}{3})^{a}\frac{\prod_{K=0}^{a-1}(N-3K)}{\prod_{K=0}^{a-1}(N-K)}$ $\leq$ $(\frac{1}{3})^{a}$ $\leq$ $(\frac{1}{3})^{\alpha/3}$. This means that the probability of Alice being tricked by Bob is exponentially small. Therefore, Protocol 1 is verifiable.
\subsection{The Presented ADBQC protocol where Alice only performs single-qubit gates}
Protocol 2 for a user Alice who can only perform single-qubit gates delegating her ADBQC to a server Bob is presented in this part. The specific steps are as follows.

(D1) Resource preparation phase: The operations in this phase are similar to the step Q1 of Protocol 1. Bob prepares the input register qubits and Alice chooses $h$ as the number of trap qubits. Note that if there are too many traps, the computational efficiency will be reduced. But if there are too few traps, the probability of detecting a malicious Bob will be small. 

(D2) Calculation phase: Protocol 2 uses the gate pattern shown in Fig. \ref{fig:a-gate-pattern} to perform ADBQC.

For the implementation of the $CZ$ gate in each gate pattern, the operation is similar as the reviewed Sueki et al.'s protocol\cite{sueki2013ancilla}. Bob should send Alice the measurement result $S^{'}$ as shown in Fig. \ref{fig:simulation-of-cz} to Alice.

\begin{figure}[h]
	\centering
	\includegraphics[width=1.0\linewidth]{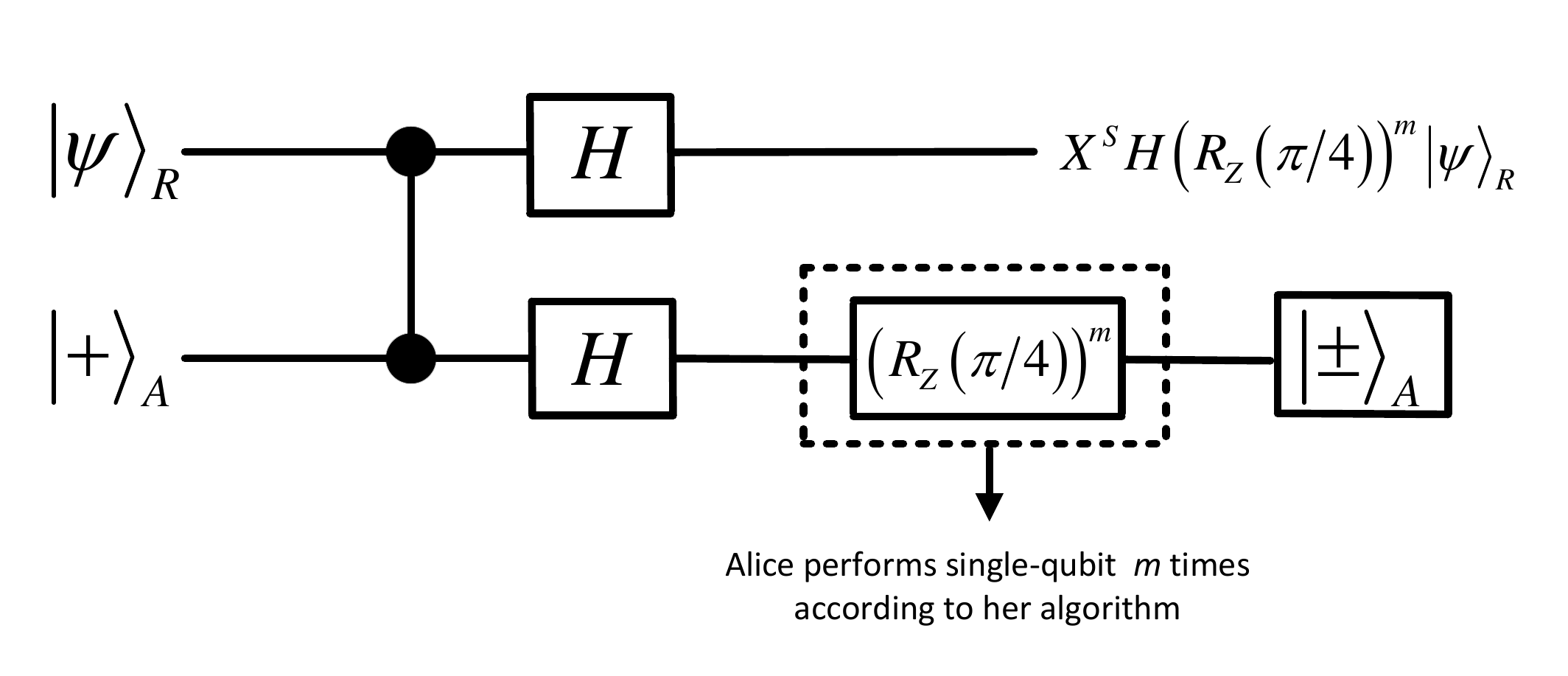}
	\caption{The figure illustrates that Alice who is only capable of performing $R_{Z}(\frac{\pi}{4})$ can achieve the operation $HR_{Z}(\frac{\pi}{4})$ with the help of Bob. The quantum operations that Alice performs in Protocol 2 are circled by the dotted lines. Alice performs $R_{Z}(\frac{\pi}{4})$ on the ancilla sent by Bob and then sends it back to Bob. After that, Bob performs $X$ basis measurement on the ancilla and informs Alice of the measurement result through the classical channel}
	\label{fig:alicesingle-qubitgate}
\end{figure}

For the implementation of $HR_{z}(\theta)$ in each gate pattern, specific operations are as follows: (1) Bob couples ancilla to register qubit and sends the coupled ancilla to Alice. (2) Alice performs $R_{z}(\frac{\pi}{4})$ gate $K$ times on ancilla according to her algorithm, and then Alice sends it back to Bob. (3) Bob measures the particle sent by Alice in the $X$ basis and sends Alice the measurement result through class channel. Its simple graphical representation is shown in Fig. \ref{fig:alicesingle-qubitgate}.

(D3) Verification phase:  At the end of Q2, let the output state be $\sigma_{q}P|\Phi\rangle$ $=$ $\sigma_{q} P(|\zeta^{'}\rangle_{r} \otimes |\lambda\rangle_{t})$, where both $|\zeta^{'}\rangle_{r}$ and $|\lambda\rangle_{t}$  are composed of $|\pm\rangle$, $|0\rangle$, and $|1\rangle$. The number of qubits contained in $|\zeta^{'}\rangle_{r}$ is $N-h$, while that in $|\lambda\rangle_{t}$ is $h$. Alice instructs Bob to perform suitable measurements on qubits of $\sigma_{q}P|\Phi\rangle$ one by one, such as $Z$ on $|0\rangle$ or $|1\rangle$ and $X$ on $|+\rangle$ or $|-\rangle$. If the error rate of the trap qubits is acceptable, Alice accepts the results of these computational register qubits. Otherwise, she rejects them.

\emph{Security analysis. }Next, blindness and verifiability of Protocol 2 are analyzed. Protocol 2 needs a bidirectional quantum channel between Alice and Bob and thus it no longer satisfies the no-signaling principle \cite{popescu1994quantum}. In fact, Bob also has no access to Alice's private information because he cannot distinguish which operations Alice did in step D2. Suppose Bob is evil and wants to capture what Alice did in step D2. He can prepare a three-particle state $|\Psi\rangle_{BBA}=a|000\rangle_{BBA}+b|001\rangle_{BBA}+c|010\rangle_{BBA}+d|100\rangle_{BBA}+e|101\rangle_{BBA}+f|110\rangle_{BBA}+g|011\rangle_{BBA}+h|111\rangle_{BBA}$ with $ \vert a \vert^{2}+\vert b \vert^{2}+\vert c \vert^{2}+\vert d \vert^{2}+\vert e \vert^{2}+\vert f \vert^{2}+\vert g \vert^{2}+\vert h \vert^{2}=1$, where subscript $A$ denotes the particles Bob sends to Alice and subscript $B$ denotes the particles retained by Bob himself. Bob retains two qubits of the quantum state and sends the remaining one to Alice, who performs $K$ times $R_{Z}(\frac{\pi}{4})$ on it and then sends it back to Bob. According to the number of $R_{Z}(\frac{\pi}{4})$ gate performed by Alice, there are eight possible states in Bob's position:
$|\Psi_{0}\rangle_{BBA} = (I\otimes I\otimes I)|\Psi\rangle_{BBA}, |\Psi_{1}\rangle_{BBA} = (I\otimes I\otimes R_{Z}(\frac{\pi}{4}))|\Psi\rangle_{BBA}, (I\otimes I\otimes R_{Z}(\frac{2\pi}{4})|\Psi\rangle_{BBA},..., |\Psi_{7}\rangle_{BBA} = (I\otimes I\otimes R_{Z}(\frac{7\pi}{4}))|\Psi\rangle_{BBA}$. If Bob wants to distinguish among these eight states by joint measurements, the eight quantum states must be orthogonal to each other, i.e., $ \vert b \vert^{2}+\vert e \vert^{2}+\vert g \vert^{2}+\vert h \vert^{2} =0$. Thus, the three-particle state prepared by Bob must be $|\Psi\rangle_{BBA} =  a|000\rangle_{BBA}+c|010\rangle_{BBA}+d|100\rangle_{BBA}+f|110\rangle_{BBA} $. Obviously, the particle that Bob assigns to Alice is not entangled with the two particles that Bob keeps, thus Bob cannot know what operations Alice performed.

\emph{Algorithm blindness of Protocol 2.}
Let $K$ be the random variable which represents the number of single-qubit gates performed by Alice, $B$ be the random variable representing the type of the POVM which Bob performs on the whole 3-qubit system,
and $M_{B}$ be the random variable which represents the result of the POVM.  Bob's knowledge about Alice's measurement angles is given by the conditional probability distribution of $K$ = $k$ given $B$ = $b$, $M_{B}$ = $m_{B}$ :
\begin{equation}
	\begin{aligned}
		P(K = k | B = b, M_{B} = m_{B}).
	\end{aligned}
\end{equation}
From Bayes' theorem, we have
\begin{equation}
	\begin{aligned}
		&P(K = k | B = b, M_{B} = m_{B})\\
		=&\frac{P(M_{B} = m_{B}, K = k, B = b)}{P(B = b, m_{B} = m_{B})}\\
		=&\frac{P(M_{B} = m_{B}, K = k, B = b)}{P(B = b, M_{B} = m_{B})}\\
		=&\frac{P(M_{B} = m_{B}|K = k, B = b)P(K = k,B = b)}{P(B = b, M_{B} = m_{B})}\\
		=&\frac{P(M_{B} = m_{B} | K = k^{'}, B = b)P(K = k^{'}, B = b)}{P(B = b, M_{B} = m_{B})}\\
		=&P(K = k^{'} | B = b, M_{B} = m_{B}).
	\end{aligned}
\end{equation}

This means that Bob cannot learn anything about the number of operator $R_{Z}(\frac{\pi}{4})$ performed by Bob.

\emph{Input Blindness of Protocol 2.}
The preparation of the input
state is included in the computational part. The input blindness of Protocol 2 is guaranteed as long as both the algorithm of Protocol 2 is blind. 

\emph{Output blindness of Protocol 2.}
Let $O$ be the random variable which represents the output of Alice's algorithm, $B$ be the random variable representing the type of the POVM which Bob performs on the whole 3-qubit system, and $M_{B}$ be the random variable which represents the result of the POVM. Bob's knowledge about the output $O$ = $o$ of Alice's algorithm is given by the conditional probability
distribution of $B$ = $b$ and $M_{B}$ = $m_{B}$:
\begin{equation}
	\begin{aligned}
		P(O = o | B = b, M_{B} = m_{B}).
	\end{aligned}
\end{equation}
From Bayes' theorem, we have the following result which shows Alice's privacy about the output is guaranteed:
\begin{equation}
	\begin{aligned}
		&P(O = o | B = b, M_{B} = m_{B})\\
		=&\frac{P(M_{B} = m_{B}, O = o, B = b)}{P(B = b, m_{B} = m_{B})}\\
		=&\frac{P(M_{B} = m_{B}, O = o, B = b)}{P(B = b, M_{B} = m_{B})}\\
		=&\frac{P(M_{B} = m_{B}|O = o, B = b)P(O = o,B = b)}{P(B = b, M_{B} = m_{B})}\\
		=&\frac{P(M_{B} = m_{B} | O = o^{'}, B = b)P(O = o^{'}, B = b)}{P(B = b, M_{B} = m_{B})}\\
		=&P(O = o^{'} | B = b, M_{B} = m_{B}).
	\end{aligned}
\end{equation}
\emph{Verifiability of Protocol 2.} Among the output register qubits, there are both the actual output qubits that are the result of the computation and the trap qubits that are used to detect Bob's honesty. We just need to select actual output qubits and trap qubits from the same set of qubits and Bob cannot distinguish between the actual output and trap qubits by the choice of measurement bases.

Suppose here the output and trap qubits are selected from $\{|0\rangle, |1\rangle, |+\rangle, |-\rangle\}$. Alice instructs Bob on the choices of the measurement bases through the classical channel. If Bob disturbs the computation to produce a different output $\rho^{'}$ instead of $\sigma_{q}P|\Phi\rangle$ or chooses different measurement bases to measure trap qubits, Bob may return incorrect measurement results at the locations of trap qubits. Without loss of generality, we can assume that the probability of each trap qubit returning an incorrect measurement is $\delta$, where $0<\delta<1$. If the number of trap qubits is $k$, the probability $p$ that Bob disrupts the computation without being dected by Alice is $\delta^{k}$. As long as $k$ is large enough, $p$ approaches zero.
\section{Conclusions }
In this paper, two ADBQC protocols have been proposed and they have different quantum capability requirements for clients. In Protocol 1, the user can perform ADBQC by only making Pauli-$X$, $Y$, and $Z$ baisis measurements. Protocol 2 requires the client to have the ability to perform single-qubit gates to achieve ADBQC with the aid of a server. The proposed ADBQC protocols have extended the exisiting ADBQC protocol in Ref. \cite{sueki2013ancilla} which just deals with the user who have the ablitiy to generate single-qubit states, and offered greater flexibility in the quantum capability requirements of clients participating in ADBQC. If the quantum capacity of the client needs to be further reduced, the double-server approach in MBQC can be considered, which can reduce the quantum capacity of the client to be totally classical. Besides, the introduction of trap qubits in the presented ADBQC protocols not only hides the scale of the algorithm, but also makes them verifiable. 

However, compared to the MBQC model, the computational efficiency in the ADBQC model is significantly lower since the number of entanglement operations that need to be performed using the ADBQC model is much higher than that are used in the MBQC model when performing the same algorithm. How to improve the computational efficiency of ADBQC protocols will be future work.

\begin{backmatter}

\newcommand{\BMCxmlcomment}[1]{}

\BMCxmlcomment{
	
	<refgrp>
	
	<bibl id="B1">
	<title><p>Quantum computational networks</p></title>
	<aug>
	<au><snm>David</snm><fnm>E. D.</fnm></au>
	</aug>
	<source>Proceedings of the Royal Society of London. A. Mathematical and
	Physical Sciences</source>
	<pubdate>1989</pubdate>
	<fpage>73</fpage>
	<lpage>-90</lpage>
	</bibl>
	
	<bibl id="B2">
	<title><p>Delegating private quantum computations</p></title>
	<aug>
	<au><snm>Broadbent</snm><fnm>A.</fnm></au>
	</aug>
	<source>Can J Phys</source>
	<pubdate>2015</pubdate>
	<volume>93</volume>
	<issue>9</issue>
	<fpage>941</fpage>
	<lpage>-946</lpage>
	</bibl>
	
	<bibl id="B3">
	<title><p>Single-server blind quantum computation with quantum circuit
	model</p></title>
	<aug>
	<au><snm>Zhang</snm><fnm>X.</fnm></au>
	<au><snm>Weng</snm><fnm>J.</fnm></au>
	<au><snm>Li</snm><fnm>X.</fnm></au>
	<au><cnm>others</cnm></au>
	</aug>
	<source>Quantum Inf Process</source>
	<publisher>Springer</publisher>
	<pubdate>2018</pubdate>
	<volume>17</volume>
	<issue>6</issue>
	<fpage>1</fpage>
	<lpage>-18</lpage>
	</bibl>
	
	<bibl id="B4">
	<title><p>A one-way quantum computer</p></title>
	<aug>
	<au><snm>Raussendorf</snm><fnm>R.</fnm></au>
	<au><snm>Briegel</snm><fnm>H. J.</fnm></au>
	</aug>
	<source>Phys Rev Lett</source>
	<publisher>APS</publisher>
	<pubdate>2001</pubdate>
	<volume>86</volume>
	<issue>22</issue>
	<fpage>5188</fpage>
	</bibl>
	
	<bibl id="B5">
	<title><p>Quantum computational universality of hypergraph states with
	{P}auli-{X} and {Z} basis measurements</p></title>
	<aug>
	<au><snm>Takeuchi</snm><fnm>Y.</fnm></au>
	<au><snm>Morimae</snm><fnm>T.</fnm></au>
	<au><snm>Hayashi</snm><fnm>M.</fnm></au>
	</aug>
	<source>Sci Rep</source>
	<pubdate>2019</pubdate>
	<volume>9</volume>
	<issue>1</issue>
	<fpage>1</fpage>
	<lpage>-14</lpage>
	</bibl>
	
	<bibl id="B6">
	<title><p>Blind quantum computing with weak coherent pulses</p></title>
	<aug>
	<au><snm>Dunjko</snm><fnm>V.</fnm></au>
	<au><snm>Kashefi</snm><fnm>E.</fnm></au>
	<au><snm>Leverrier</snm><fnm>A.</fnm></au>
	</aug>
	<source>Phys Rev Lett</source>
	<publisher>APS</publisher>
	<pubdate>2012</pubdate>
	<volume>108</volume>
	<issue>20</issue>
	<fpage>200502</fpage>
	</bibl>
	
	<bibl id="B7">
	<title><p>Blind topological measurement-based quantum computation</p></title>
	<aug>
	<au><snm>Morimae</snm><fnm>T.</fnm></au>
	<au><snm>Fujii</snm><fnm>K.</fnm></au>
	</aug>
	<source>Nat Commun</source>
	<publisher>Nature Publishing Group</publisher>
	<pubdate>2012</pubdate>
	<volume>3</volume>
	<issue>1</issue>
	<fpage>1</fpage>
	<lpage>-6</lpage>
	</bibl>
	
	<bibl id="B8">
	<title><p>Blind quantum computation protocol in which Alice only makes
	measurements</p></title>
	<aug>
	<au><snm>Tomoyuki</snm><fnm>T.</fnm></au>
	<au><snm>Fujii</snm><fnm>K.</fnm></au>
	</aug>
	<source>Phys Rev A</source>
	<publisher>APS</publisher>
	<pubdate>2013</pubdate>
	<volume>87</volume>
	<issue>5</issue>
	<fpage>050301</fpage>
	</bibl>
	
	<bibl id="B9">
	<title><p>Secure entanglement distillation for double-server blind quantum
	computation</p></title>
	<aug>
	<au><cnm>{T. Tomoyuki and K. Fujii}</cnm></au>
	</aug>
	<source>Phys Rev Lett</source>
	<pubdate>2013</pubdate>
	<volume>111</volume>
	<issue>2</issue>
	<fpage>020502</fpage>
	</bibl>
	
	<bibl id="B10">
	<title><p>Triple-server blind quantum computation using entanglement
	swapping</p></title>
	<aug>
	<au><snm>Li</snm><fnm>Q.</fnm></au>
	<au><snm>Chan</snm><fnm>W. H.</fnm></au>
	<au><snm>Wu</snm><fnm>C.</fnm></au>
	<au><cnm>others</cnm></au>
	</aug>
	<source>Phys Rev A</source>
	<publisher>APS</publisher>
	<pubdate>2014</pubdate>
	<volume>89</volume>
	<issue>4</issue>
	<fpage>040302</fpage>
	</bibl>
	
	<bibl id="B11">
	<title><p>Deterministic entanglement distillation for secure double-server
	blind quantum computation</p></title>
	<aug>
	<au><snm>Sheng</snm><fnm>Y. B.</fnm></au>
	<au><snm>Zhou</snm><fnm>L.</fnm></au>
	</aug>
	<source>Sci Rep</source>
	<pubdate>2015</pubdate>
	<volume>5</volume>
	<issue>1</issue>
	<fpage>1</fpage>
	<lpage>-5</lpage>
	</bibl>
	
	<bibl id="B12">
	<title><p>Hierarchy of universal entanglement in 2{D} measurement-based
	quantum computation</p></title>
	<aug>
	<au><snm>Miller</snm><fnm>J.</fnm></au>
	<au><snm>Miyake</snm><fnm>A.</fnm></au>
	</aug>
	<source>npj Quantum Inform</source>
	<publisher>Nature Publishing Group</publisher>
	<pubdate>2016</pubdate>
	<volume>2</volume>
	<issue>1</issue>
	<fpage>1</fpage>
	<lpage>-6</lpage>
	</bibl>
	
	<bibl id="B13">
	<title><p>Ancilla-driven universal quantum computation</p></title>
	<aug>
	<au><snm>Anders</snm><fnm>J.</fnm></au>
	<au><snm>Oi</snm><fnm>D. K. L.</fnm></au>
	<au><snm>Kashefi</snm><fnm>E.</fnm></au>
	<au><snm>Browne</snm><fnm>D. E.</fnm></au>
	<au><cnm>others</cnm></au>
	</aug>
	<source>Phys Rev A</source>
	<pubdate>2010</pubdate>
	<volume>82</volume>
	<issue>2</issue>
	<fpage>020301</fpage>
	</bibl>
	
	<bibl id="B14">
	<title><p>Entanglement of atoms via cold controlled collisions</p></title>
	<aug>
	<au><snm>Jaksch</snm><fnm>D.</fnm></au>
	<au><snm>Briegel</snm><fnm>H. J.</fnm></au>
	<au><snm>Cirac</snm><fnm>J. I.</fnm></au>
	<au><cnm>others</cnm></au>
	</aug>
	<source>Phys Rev Lett</source>
	<pubdate>1999</pubdate>
	<volume>82</volume>
	<issue>9</issue>
	<fpage>1975</fpage>
	</bibl>
	
	<bibl id="B15">
	<title><p>Coupling superconducting qubits via a cavity bus</p></title>
	<aug>
	<au><snm>Majer</snm><fnm>J.</fnm></au>
	<au><snm>Chow</snm><fnm>J. M.</fnm></au>
	<au><snm>Gambetta</snm><fnm>J. M.</fnm></au>
	<au><cnm>others</cnm></au>
	</aug>
	<source>Nature</source>
	<pubdate>2007</pubdate>
	<volume>449</volume>
	<issue>7161</issue>
	<fpage>443</fpage>
	<lpage>-447</lpage>
	</bibl>
	
	<bibl id="B16">
	<title><p>Frequency comparison of two high-accuracy Al optical
	clocks</p></title>
	<aug>
	<au><snm>Chou</snm><fnm>C.</fnm></au>
	<au><snm>Hume</snm><fnm>D. B.</fnm></au>
	<au><snm>Koelemeij</snm><fnm>J. C.</fnm></au>
	<au><cnm>others</cnm></au>
	</aug>
	<source>Phys Rev Lett</source>
	<pubdate>2010</pubdate>
	<volume>104</volume>
	<issue>7</issue>
	<fpage>070802</fpage>
	</bibl>
	
	<bibl id="B17">
	<title><p>Effect on ion-trap quantum computers from the quantum nature of the
	driving field</p></title>
	<aug>
	<au><snm>Yang</snm><fnm>B.</fnm></au>
	<au><snm>Yang</snm><fnm>L.</fnm></au>
	</aug>
	<source>Sci China Inf Sci</source>
	<pubdate>2020</pubdate>
	<volume>63</volume>
	<issue>10</issue>
	<fpage>202501</fpage>
	</bibl>
	
	<bibl id="B18">
	<title><p>A quantum key distribution protocol for rapid denial of service
	detection</p></title>
	<aug>
	<au><snm>Price</snm><fnm>AB</fnm></au>
	<au><snm>Rarity</snm><fnm>JG</fnm></au>
	<au><snm>Erven</snm><fnm>C</fnm></au>
	</aug>
	<source>EPJ Quantum Technology</source>
	<publisher>Springer Berlin Heidelberg</publisher>
	<pubdate>2020</pubdate>
	<volume>7</volume>
	<issue>1</issue>
	<fpage>8</fpage>
	</bibl>
	
	<bibl id="B19">
	<title><p>Quantum key distribution based on single-particle and EPR
	entanglement</p></title>
	<aug>
	<au><snm>Li</snm><fnm>L.</fnm></au>
	<au><snm>Li</snm><fnm>J.</fnm></au>
	<au><snm>Yan</snm><fnm>C.</fnm></au>
	<au><cnm>others</cnm></au>
	</aug>
	<source>Sci China Inf Sci</source>
	<pubdate>2020</pubdate>
	<volume>63</volume>
	<issue>6</issue>
	<fpage>169501</fpage>
	</bibl>
	
	<bibl id="B20">
	<title><p>Blind quantum computation with identity authentication</p></title>
	<aug>
	<au><snm>Li</snm><fnm>Q.</fnm></au>
	<au><snm>Li</snm><fnm>Z.</fnm></au>
	<au><snm>Chan</snm><fnm>W. H.</fnm></au>
	<au><cnm>others</cnm></au>
	</aug>
	<source>Phys Lett A</source>
	<pubdate>2018</pubdate>
	<volume>382</volume>
	<issue>14</issue>
	<fpage>938</fpage>
	<lpage>-941</lpage>
	</bibl>
	
	<bibl id="B21">
	<title><p>An efficient anti-quantum lattice-based blind signature for
	blockchain-enabled systems</p></title>
	<aug>
	<au><snm>Li</snm><fnm>C.</fnm></au>
	<au><snm>Tian</snm><fnm>Y.</fnm></au>
	<au><snm>Chen</snm><fnm>X.</fnm></au>
	<au><cnm>others</cnm></au>
	</aug>
	<source>Inf Sci</source>
	<pubdate>2021</pubdate>
	<volume>546</volume>
	<fpage>253</fpage>
	<lpage>264</lpage>
	</bibl>
	
	<bibl id="B22">
	<title><p>Hierarchical quantum secret sharing based on special
	high-dimensional entangled state</p></title>
	<aug>
	<au><snm>Qin</snm><fnm>H.</fnm></au>
	<au><snm>Tang</snm><fnm>W. K.</fnm></au>
	<au><snm>Tso</snm><fnm>R.</fnm></au>
	</aug>
	<source>IEEE J SEL TOP QUANT</source>
	<pubdate>2020</pubdate>
	<volume>26</volume>
	<issue>3</issue>
	<fpage>1</fpage>
	<lpage>-6</lpage>
	</bibl>
	
	<bibl id="B23">
	<title><p>Secure assisted quantum computation</p></title>
	<aug>
	<au><snm>Childs</snm><fnm>A. M.</fnm></au>
	</aug>
	<source>Quantum Inf Comput</source>
	<pubdate>2005</pubdate>
	<volume>5</volume>
	<issue>6</issue>
	<fpage>456</fpage>
	<lpage>-466</lpage>
	</bibl>
	
	<bibl id="B24">
	<title><p>Universal blind quantum computation</p></title>
	<aug>
	<au><snm>{Broadbent}</snm><fnm>A.</fnm></au>
	<au><snm>{F.Fitzsimons}</snm><fnm>J.</fnm></au>
	<au><snm>{Kashefi}</snm><fnm>E.</fnm></au>
	</aug>
	<source>Proceedings of the 50th Annual IEEE Symposium on Foundations of
	Computer Science</source>
	<pubdate>2009</pubdate>
	<fpage>517</fpage>
	<lpage>-526</lpage>
	</bibl>
	
	<bibl id="B25">
	<title><p>Robustness and device independence of verifiable blind quantum
	computing</p></title>
	<aug>
	<au><snm>Gheorghiu</snm><fnm>A.</fnm></au>
	<au><snm>Kashefi</snm><fnm>E.</fnm></au>
	<au><snm>Wallden</snm><fnm>P.</fnm></au>
	</aug>
	<source>New J Phys</source>
	<pubdate>2015</pubdate>
	<volume>17</volume>
	<issue>8</issue>
	<fpage>083040</fpage>
	</bibl>
	
	<bibl id="B26">
	<title><p>Verifiable measurement-only blind quantum computing with stabilizer
	testing</p></title>
	<aug>
	<au><snm>Hayashi</snm><fnm>M.</fnm></au>
	<au><snm>Morimae</snm><fnm>T.</fnm></au>
	</aug>
	<source>Phys Rev lett</source>
	<pubdate>2015</pubdate>
	<volume>115</volume>
	<issue>22</issue>
	<fpage>220502</fpage>
	</bibl>
	
	<bibl id="B27">
	<title><p>Unconditionally verifiable blind quantum computation</p></title>
	<aug>
	<au><snm>Fitzsimons</snm><fnm>J. F.</fnm></au>
	<au><snm>Kashefi</snm><fnm>E.</fnm></au>
	</aug>
	<source>Phys Rev A</source>
	<pubdate>2017</pubdate>
	<volume>96</volume>
	<issue>1</issue>
	<fpage>012303</fpage>
	</bibl>
	
	<bibl id="B28">
	<title><p>Measurement-only verifiable blind quantum computing with quantum
	input verification</p></title>
	<aug>
	<au><snm>Morimae</snm><fnm>T.</fnm></au>
	</aug>
	<source>Phys Rev A</source>
	<pubdate>2016</pubdate>
	<volume>94</volume>
	<issue>4</issue>
	<fpage>042301</fpage>
	</bibl>
	
	<bibl id="B29">
	<title><p>Universal blind quantum computation for hybrid system</p></title>
	<aug>
	<au><snm>Huang</snm><fnm>H. L.</fnm></au>
	<au><snm>Bao</snm><fnm>W. S.</fnm></au>
	<au><snm>Li</snm><fnm>T.</fnm></au>
	<au><cnm>others</cnm></au>
	</aug>
	<source>Quantum Inf Process</source>
	<publisher>Springer</publisher>
	<pubdate>2017</pubdate>
	<volume>16</volume>
	<issue>8</issue>
	<fpage>1</fpage>
	<lpage>-8</lpage>
	</bibl>
	
	<bibl id="B30">
	<title><p>Self-guaranteed measurement-based quantum computation</p></title>
	<aug>
	<au><snm>Hayashi</snm><fnm>M.</fnm></au>
	<au><snm>Hajdu{\v{s}}ek</snm><fnm>M.</fnm></au>
	</aug>
	<source>Phys Rev A</source>
	<publisher>APS</publisher>
	<pubdate>2018</pubdate>
	<volume>97</volume>
	<issue>5</issue>
	<fpage>052308</fpage>
	</bibl>
	
	<bibl id="B31">
	<title><p>Verification for measurement-only blind quantum
	computing</p></title>
	<aug>
	<au><snm>Morimae</snm><fnm>T.</fnm></au>
	</aug>
	<source>Phys Rev A</source>
	<publisher>APS</publisher>
	<pubdate>2014</pubdate>
	<volume>89</volume>
	<issue>6</issue>
	<fpage>060302</fpage>
	</bibl>
	
	<bibl id="B32">
	<title><p>Verification of many-qubit states</p></title>
	<aug>
	<au><snm>Takeuchi</snm><fnm>Y.</fnm></au>
	<au><snm>Morimae</snm><fnm>T.</fnm></au>
	</aug>
	<source>Phys Rev X</source>
	<publisher>APS</publisher>
	<pubdate>2018</pubdate>
	<volume>8</volume>
	<issue>2</issue>
	<fpage>021060</fpage>
	</bibl>
	
	<bibl id="B33">
	<title><p>Arbitrable blind quantum computation</p></title>
	<aug>
	<au><snm>G</snm><fnm>S</fnm></au>
	<au><snm>T</snm><fnm>K</fnm></au>
	<au><snm>T</snm><fnm>M</fnm></au>
	</aug>
	<source>Quantum Inform Process</source>
	<publisher>Springer</publisher>
	<pubdate>2019</pubdate>
	<volume>18</volume>
	<issue>12</issue>
	<fpage>1</fpage>
	<lpage>-8</lpage>
	</bibl>
	
	<bibl id="B34">
	<title><p>Multi-party blind quantum computation protocol with mutual
	authentication in network</p></title>
	<aug>
	<au><snm>Shan</snm><fnm>R. T.</fnm></au>
	<au><snm>Chen</snm><fnm>X.</fnm></au>
	<au><snm>Yuan</snm><fnm>K. G.</fnm></au>
	</aug>
	<source>Sci China Inf Sci</source>
	<publisher>Science China Press</publisher>
	<pubdate>2021</pubdate>
	<volume>64</volume>
	<issue>6</issue>
	<fpage>162302</fpage>
	</bibl>
	
	<bibl id="B35">
	<title><p>A hybrid universal blind quantum computation</p></title>
	<aug>
	<au><snm>Zhang</snm><fnm>X.</fnm></au>
	<au><snm>Luo</snm><fnm>W.</fnm></au>
	<au><snm>Zeng</snm><fnm>G.</fnm></au>
	<au><cnm>others</cnm></au>
	</aug>
	<source>Inf Sci</source>
	<pubdate>2019</pubdate>
	<volume>498</volume>
	<fpage>135</fpage>
	<lpage>143</lpage>
	</bibl>
	
	<bibl id="B36">
	<title><p>Demonstration of blind quantum computing</p></title>
	<aug>
	<au><snm>Barz</snm><fnm>S.</fnm></au>
	<au><snm>Kashefi</snm><fnm>E.</fnm></au>
	<au><snm>Broadbent</snm><fnm>A.</fnm></au>
	<au><cnm>others</cnm></au>
	</aug>
	<source>Science</source>
	<publisher>American Association for the Advancement of Science</publisher>
	<pubdate>2012</pubdate>
	<volume>335</volume>
	<issue>6066</issue>
	<fpage>303</fpage>
	<lpage>-308</lpage>
	</bibl>
	
	<bibl id="B37">
	<title><p>Experimental verification of quantum computation</p></title>
	<aug>
	<au><snm>Barz</snm><fnm>S.</fnm></au>
	<au><snm>Fitzsimons</snm><fnm>J. F.</fnm></au>
	<au><snm>Kashefi</snm><fnm>E.</fnm></au>
	<au><cnm>others</cnm></au>
	</aug>
	<source>Nat Phys</source>
	<publisher>Nature Publishing Group</publisher>
	<pubdate>2013</pubdate>
	<volume>9</volume>
	<issue>11</issue>
	<fpage>727</fpage>
	<lpage>-731</lpage>
	</bibl>
	
	<bibl id="B38">
	<title><p>Blind quantum computation where a user only performs single-qubit
	gates</p></title>
	<aug>
	<au><snm>Li</snm><fnm>Q.</fnm></au>
	<au><snm>Liu</snm><fnm>C.</fnm></au>
	<au><snm>Peng</snm><fnm>Y.</fnm></au>
	<au><cnm>others</cnm></au>
	</aug>
	<source>Opt Laser Technol</source>
	<publisher>Elsevier</publisher>
	<pubdate>2021</pubdate>
	<volume>142</volume>
	<fpage>107190</fpage>
	</bibl>
	
	<bibl id="B39">
	<title><p>Ancilla-driven universal blind quantum computation</p></title>
	<aug>
	<au><snm>Sueki</snm><fnm>T.</fnm></au>
	<au><snm>Koshiba</snm><fnm>T.</fnm></au>
	<au><snm>Morimae</snm><fnm>T.</fnm></au>
	</aug>
	<source>Phys Rev A</source>
	<publisher>APS</publisher>
	<pubdate>2013</pubdate>
	<volume>87</volume>
	<issue>6</issue>
	<fpage>060301</fpage>
	</bibl>
	
	<bibl id="B40">
	<title><p>Minimal ancilla mediated quantum computation</p></title>
	<aug>
	<au><snm>Proctor</snm><fnm>TJ</fnm></au>
	<au><snm>Kendon</snm><fnm>V</fnm></au>
	</aug>
	<source>EPJ quantum technology</source>
	<publisher>Springer</publisher>
	<pubdate>2014</pubdate>
	<volume>1</volume>
	<issue>1</issue>
	<fpage>1</fpage>
	<lpage>-11</lpage>
	</bibl>
	
	<bibl id="B41">
	<title><p>Ancilla-driven quantum computation with twisted graph
	states</p></title>
	<aug>
	<au><snm>Anders</snm><fnm>J.</fnm></au>
	<au><snm>Andersson</snm><fnm>E.</fnm></au>
	<au><snm>Browne</snm><fnm>D. E.</fnm></au>
	<au><cnm>others</cnm></au>
	</aug>
	<source>Theor Comput Sci</source>
	<publisher>Elsevier</publisher>
	<pubdate>2012</pubdate>
	<volume>430</volume>
	<fpage>51</fpage>
	<lpage>-72</lpage>
	</bibl>
	
	<bibl id="B42">
	<title><p>Time optimal control in spin systems</p></title>
	<aug>
	<au><snm>Khaneja</snm><fnm>N.</fnm></au>
	<au><snm>Brockett</snm><fnm>R.</fnm></au>
	<au><snm>Glaser</snm><fnm>S. J.</fnm></au>
	</aug>
	<source>Phys Rev A</source>
	<publisher>APS</publisher>
	<pubdate>2001</pubdate>
	<volume>63</volume>
	<issue>3</issue>
	<fpage>032308</fpage>
	</bibl>
	
	<bibl id="B43">
	<title><p>Quantum nonlocality as an axiom</p></title>
	<aug>
	<au><snm>Popescu</snm><fnm>S.</fnm></au>
	<au><snm>Rohrlich</snm><fnm>D.</fnm></au>
	</aug>
	<source>Found Phys</source>
	<publisher>Springer</publisher>
	<pubdate>1994</pubdate>
	<volume>24</volume>
	<issue>3</issue>
	<fpage>379</fpage>
	<lpage>-385</lpage>
	</bibl>
	
	</refgrp>
} 

\end{backmatter}
\end{document}